\documentclass[aps,prb,twocolumn,superscriptaddress,showpacs]{revtex4}

\usepackage{graphicx}
\usepackage{dcolumn}
\usepackage{bm}

\begin{document}
\title{Heat transport of the quasi-one-dimensional alternating spin
chain material (CH$_3$)$_2$NH$_2$CuCl$_3$}

\author{L. M. Chen}
\affiliation{Hefei National Laboratory for Physical Sciences at
Microscale, University of Science and Technology of China, Hefei,
Anhui 230026, People's Republic of China}

\affiliation{Department of Physics, University of Science and
Technology of China, Hefei, Anhui 230026, People's Republic of
China}

\author{X. M. Wang}
\affiliation{Hefei National Laboratory for Physical Sciences at
Microscale, University of Science and Technology of China, Hefei,
Anhui 230026, People's Republic of China}

\author{W. P. Ke}
\affiliation{Hefei National Laboratory for Physical Sciences at
Microscale, University of Science and Technology of China, Hefei,
Anhui 230026, People's Republic of China}

\author{Z. Y. Zhao}
\affiliation{Hefei National Laboratory for Physical Sciences at
Microscale, University of Science and Technology of China, Hefei,
Anhui 230026, People's Republic of China}

\author{X. G. Liu}
\affiliation{Hefei National Laboratory for Physical Sciences at
Microscale, University of Science and Technology of China, Hefei,
Anhui 230026, People's Republic of China}

\author{C. Fan}
\affiliation{Hefei National Laboratory for Physical Sciences at
Microscale, University of Science and Technology of China, Hefei,
Anhui 230026, People's Republic of China}

\author{Q. J. Li}
\affiliation{Hefei National Laboratory for Physical Sciences at
Microscale, University of Science and Technology of China, Hefei,
Anhui 230026, People's Republic of China}

\author{X. Zhao}
\affiliation{School of Physical Sciences, University of Science
and Technology of China, Hefei, Anhui 230026, People's Republic of
China}

\author{X. F. Sun}
\email{xfsun@ustc.edu.cn}
\affiliation{Hefei National Laboratory for
Physical Sciences at Microscale, University of Science and
Technology of China, Hefei, Anhui 230026, People's Republic of
China}

\date{\today}

\begin{abstract}

We report a study of the low-temperature heat transport in the
quasi-one-dimensional $S$ = 1/2 alternating
antiferromagnetic-ferromagnetic chain compound
(CH$_3$)$_2$NH$_2$CuCl$_3$. Both the temperature and
magnetic-field dependencies of thermal conductivity are very
complicated, pointing to the important role of spin excitations.
It is found that magnetic excitations act mainly as the phonon
scatterers in a broad temperature region from 0.3 to 30 K. In
magnetic fields, the thermal conductivity show drastic changes,
particularly at the field-induced transitions from the low-field
N\'{e}el state to the spin-gapped state, the field-induced
magnetic ordered state, and the spin polarized state. In high
fields, the phonon conductivity is significantly enhanced because
of the weakening of spin fluctuations.

\end{abstract}


\maketitle

\section{Introduction}

Low-dimensional or frustrated quantum magnets were revealed to
exhibit exotic ground states, magnetic excitations, and quantum
phase transitions (QPTs).\cite{Sachdev, Balents} For a particular
case of the spin-gapped antiferromagnets, the external magnetic
field can close the gap in the spectrum, which results in a QPT
between a low-field disordered paramagnetic phase and a high-field
long-range ordered one. An intriguing finding is that this ordered
phase can be approximately described as a Bose-Einstein
condensation (BEC) of magnons.\cite{BEC_Review} Heat transport of
low-dimensional quantum magnets has recently received an intensive
research interests because it is very useful to probe the nature
of magnetic excitations and the field-induced QPTs.\cite{Brenig1,
Hess1, Sologubenko1, Ando, Sun_DTN} In particular, a large spin
thermal conductivity in spin-chain and spin-ladder systems has
been theoretically predicted and experimentally confirmed in such
compounds as SrCuO$_2$, Sr$_2$CuO$_3$, CaCu$_2$O$_3$,
Sr$_{14}$Cu$_{24}$O$_{41}$, etc.\cite{Sologubenko2, Hess2, Hess3,
Kawamata} Most of these materials have simple spin structure and
strong exchange coupling, which are necessary for producing
high-velocity and long-range-correlated spin excitations, while
the low dimensionality strongly enhances the quantum fluctuations
and ensures a large population of spin excitations. However, the
ground states of these materials usually have weak response to the
magnetic field because the laboratory fields are too small,
compared to the exchange energy. They are, in this sense, not
suitable for studying the field-induced QPTs and the associated
physics of magnetic excitations. Apparently, some organic magnetic
materials have obvious advantages since their larger crystal unit
cells and atom distances lead to much weaker exchange coupling of
magnetic ions. Several materials have been studied to reveal how
the heat transport behaves at the field-induced
QPTs.\cite{Sun_DTN, Sologubenko3, Sologubenko4, Sologubenko5} It
seems that the spin excitations are often playing a role of
scattering phonons and therefore strongly suppress the thermal
conductivity at the phase transitions. One alternative case is
NiCl$_2$-$4$SC(NH$_2$)$_2$ (DTN), in which the thermal
conductivity is significantly enhanced in its BEC
state.\cite{Sun_DTN, Kohama_DTN} Therefore, the role of magnetic
excitations in the heat transport at QPTs still needs to be
carefully studied.

(CH$_3$)$_2$NH$_2$CuCl$_3$ (Dimethylammonium copper $\amalg$
chloride, also known as DMACuCl$_3$ or MCCL) is an $S$ = 1/2
alternating antiferromagnetic-ferromagnetic (AFM-FM) dimer-chain
system with weak inter-dimer AF coupling.\cite{Ajiro, Willett,
Inagaki1, Yoshida1, Furukawa, Stone1, Stone2, Yoshida2, Inagaki2}
MCCL crystallizes in the monoclinic structure (space group $C2/c$)
with room-temperature lattice constants $a$ = 17.45 \AA, $b$ =
8.63 \AA, $c$ = 11.97 \AA, and $\beta$ =
125.41$^\circ$.\cite{Willett, note} The crystal structure consists
of Cu-Cl-Cu bonded chains along the $c$ axis with
Cu-Cl$\cdots$Cl-Cu contacts along the $a$ axis. These Cu-halide
planes are separated from one another along the $b$ axis by methyl
groups, so the magnetic coupling is only expected in the $ac$
plane. The basic features of the spin structure are illustrated in
Fig. 1(a). All the spin interactions $J_1$, $J_3$, $J_A$ and $J_B$
are AFM except that $J_2$ is FM. The calculations using the
diagonalization method found the relations $J_B < J_A < J_1 <
|J_2|$, and $J_3$ is nearly zero.\cite{Willett} The magnetic
structure can be regarded as an alternating predominant AFM
dimer(AF$_1$) and FM dimer (F$_2$) with the practically equal
magnitude of the intra-dimer interactions $J_1$ and $J_2$,
respectively, which are linked by weak AFM inter-dimer interaction
$J_3$. It can be modelled as
-F$_2$-AF$_3$-AF$_1$-AF$_3$-F$_2$-.\cite{Yoshida2} This rather
complicated magnetic structure results in interesting magnetic
properties and multiple phase transitions, as shown in Fig.
1(b).\cite{Inagaki1, Yoshida1, Furukawa, Yoshida2, Inagaki2} The
ground state of MCCL in the absence of the magnetic field can be
viewed as a long-range ordered chain in which alternative $S$ = 1
and $S$ = 0 spins coupled by an intervening weak interaction. A
spontaneous AFM order is formed below the N\'{e}el temperature
$T_N$ = 0.9 K, but it is stable only in the low magnetic fields
($H < H_{c1}$, $H_{c1}$ = 2 T at zero-$T$ limit), and then becomes
a field-induced gapped state in $H_{c1}< H < H_{c2}$ ($H_{c2}$ =
3.5 T at zero-$T$ limit) and finally turns into a field-induced
magnetic order (FIMO) phase at $H > H_{c2}$. The rich magnetic
phenomena of MCCL make it a good material to study the low-$T$
heat transport and its relationship with the QPTs.

\begin{figure}
\includegraphics[clip,width=8.5cm]{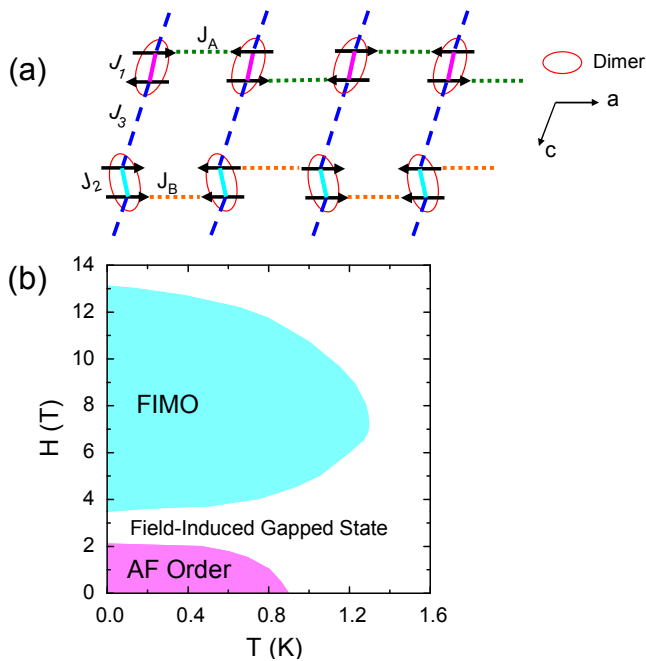}
\caption{(Color online) (a) A schematic plot of the spin structure
of MCCL. There are two types of dimers along the $a$ axis: (i) the
AFM dimers with the intra-dimer interaction $J_1$ and the
neighboring AFM interaction $J_A$; (ii) the FM dimers with the
intra-dimer interaction $J_2$ and the neighboring AFM interaction
$J_B$. Spin chain (along the $c$ axis) are contacted by
alternating AFM dimers and FM dimers through AFM interaction $J_3$
between two types of dimers. (b) The $H-T$ phase diagram of MCCL
obtained from the former experiments.\cite{Inagaki1, Yoshida1,
Furukawa, Yoshida2, Inagaki2} The magenta area and the cyan one
represent the low-field spontaneous order state and the high-field
induced magnetic order state, respectively, with the field-induced
gapped state between them.}
\end{figure}

In this work, we study the detailed temperature- and
field-dependencies of low-$T$ thermal conductivity of high-quality
MCCL single crystals. It is found that the magnetic excitations do
not transport heat directly and there is strong scattering between
magnetic excitations and phonons at zero or low fields. The phonon
thermal conductivity shows drastic changes across all the phase
transitions mentioned above. In high-field spin-polarized state,
the magnetic scattering is strongly weakened and the phonon
conductivity is significantly increased in a very broad
temperature region.

\section{Experiments}

MCCL single crystals are grown using a slow evaporation
method.\cite{Chen} Shining crystals with the typical size of
(3--6) $\times$ (1--3) $\times$ (1--2) mm$^3$, with the $bc$
crystallinity plane the largest naturally formed surface, are
selected and polished into a parallelepiped shape for the specific
heat and thermal conductivity measurements. The specific heat is
measured by the relaxation method in the temperature range from
0.4 to 10 K using a commercial physical property measurement
system (PPMS, Quantum Design). The temperature and magnetic field
dependencies of the thermal conductivity are measured using a
conventional steady-state technique in a $^3$He refrigerator and a
14 T magnet.\cite{Sun_DTN, Wang_HMO, Zhao_NCO} The heat flow is
along the $bc$ plane, in which the spin chains are included. Note
that the MCCL crystals are somewhat fragile, so the trials to make
well-shaped samples with other orientations are not successful. In
both specific-heat and thermal-conductivity measurements, the
magnetic field is applied perpendicular to the $bc$ plane.

\section{Results and Discussion}

\begin{figure}
\includegraphics[clip,width=6.0cm]{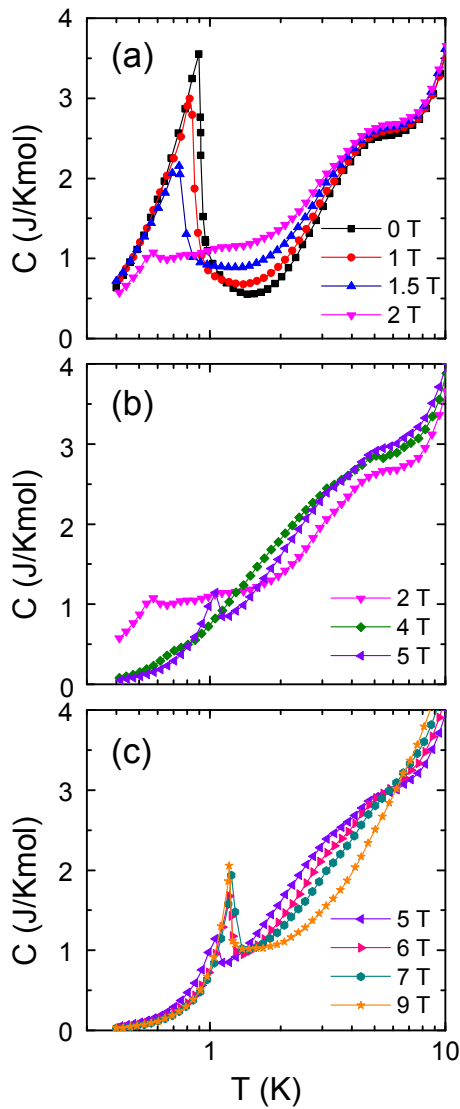}
\caption{(Color online) Specific heat of MCCL as a function of
temperature for the magnetic field applied perpendicular to the
$bc$ plane.}
\end{figure}

The low-temperature specific heat data are measured for verifying
the $H-T$ phase diagram of our MCCL crystals. Figure 2 shows the
data with magnetic field from 0 to 9 T. The main features of the
zero-field $C(T)$ curve include a narrow and sharp peak at 0.89 K
and a ``shoulder" at about 5 K, as shown in Fig. 2(a). With
increasing the field, the low-$T$ peak shifts to lower temperature
with the peak amplitude decreased and finally evolves to a very
small peak at 0.57 K when the magnetic field is increased to 2 T.
This low-$T$ peak is clearly due to the spontaneous N\'{e}el
transition, which can be suppressed by the magnetic field. With
increasing magnetic field further, this low-$T$ peak completely
disappears and a new peak shows up. It can be seen from Figs. 2(b)
and 2(c) that the new small peak appears at 1.06 K in 4-T field
and it becomes bigger and sharper and shifts to a bit higher
temperature with increasing magnetic field, in contrast to the
field dependence of the low-field peak. More exactly, the peak
temperature is highest at 7 T. Apparently, this new peak appeared
in higher fields has another origin, which is known to be the
transition of the FIMO phase. The phase boundaries determined from
the present specific-heat data have good consistency with the
former result (see Fig. 1(b)).\cite{Yoshida1}

\begin{figure}
\includegraphics[clip,width=8.0cm]{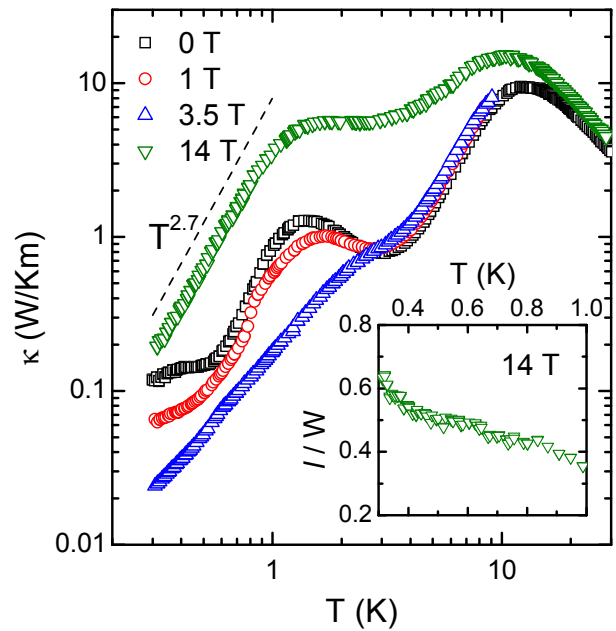}
\caption{(Color online) Temperature dependencies of the thermal
conductivity of MCCL single crystal in the zero field and several
different magnetic fields up to 14 T, which are applied
perpendicular to the $bc$ plane. The dashed line indicates a
$T^{2.7}$ dependence of $\kappa$ at subKelvin temperatures. The
inset shows the temperature dependence of the phonon mean free
path $l$ divided by the averaged sample width $W$ in 14 T magnetic
field.}
\end{figure}

Figure 3 shows the temperature dependencies of $\kappa$ in zero
field and several magnetic fields up to 14 T. In the zero field,
the temperature dependence of $\kappa$ is rather complicated. The
peak at $\sim$ 12 K is likely the phonon peak as the insulators
commonly have.\cite{Berman} At lower temperatures, there appear
two ``diplike" features in the $\kappa(T)$ curve at $\sim$ 3 K and
$\sim$ 0.6 K, respectively. In general, the possible reasons of
these behaviors in magnetic materials could be either the strong
phonon scattering by critical spin fluctuations at some magnetic
phase transitions or the phonon resonant scattering by some
magnetic impurities or lattice defects.\cite{Berman, Sun_GBCO,
Wang_HMO} The underlying mechanism can usually be judged from the
magnetic-field dependence of the diplike features of $\kappa(T)$.
As can be seen from Fig. 3 that applying magnetic field induces
drastic changes in the magnitude and the temperature dependence of
$\kappa$. As far as the lower-$T$ dip is concerned, it is found
that applying magnetic fields up to 3.5 T suppresses the
very-low-$T$ $\kappa$ so strongly that the dip is markedly
weakened in 1 T and completely disappeared in 3.5 T. Apparently,
this dip is related to the spontaneous AF ordering, which is known
to be suppressed with applying magnetic field. More exactly, in
zero field, the phonon scattering by magnetic excitations or spin
fluctuations is strong at high temperatures; upon lowering
temperature to the spontaneous AF state, the spin fluctuations are
significantly weakened and the phonon conductivity can be
increased. Thus, a diplike feature is produced. However, applying
magnetic field (up to several Tesla) can suppress the magnetic
order and somehow enhance the spin fluctuations and their
scattering on phonons. In passing, the temperature of the low-$T$
dip locates at 0.6 K, a bit lower than the AF transition
temperature 0.9 K from the specific-heat data. Similar phenomenon
was also observed in some other magnetic material.\cite{Wang_HMO}

The 3-K dip is also weakened in magnetic fields although its
position is nearly independent on the field. From the position of
this dip, it is possible to be attributed to the phonon resonant
scattering by the energy gap in the spin spectrum. The neutron
scattering measurements have revealed that in the paramagnetic
phase there are two magnon branches along the $bc$ plane; one is
dispersive and the other one is dispersionless with gaps of 0.95
and 1.6 meV, respectively.\cite{Stone1, Stone2, note} In this
situation, phonons with 0.95 meV ($\sim$ 11 K) can be resonantly
scattered by producing magnetic excitations. Since the phonon
conductivity spectrum $\kappa(\omega)$ has a (broad) maximum at
$\sim$ 3.8$k_{B}T$,\cite{Berman, Sun_GBCO} this magnetic
scattering on phonons is therefore the strongest at $\sim$ 3 K,
which agrees well with the position of the dip. The weak
dependence of dip position on the magnetic field suggests that the
gap of magnetic spectra is insensitive to the field at the
paramagnetic state.

In magnetic field as high as 14 T, the thermal conductivity show
very drastic changes. First, the magnitude of $\kappa$ is always
enhanced in a broad temperature region from 0.3 to 30 K and the
enhancement is rather large. Second, the 3-K dip in zero field
evolutes into a shoulderlike feature, suggesting that the phonon
resonant scattering is still active at 14 T. Third, at subKelvin
temperatures $\kappa(T)$ show an approximate $T^{2.7}$ dependence,
which indicates that the phonon boundary scattering is approached.
It is useful to calculate the mean free path of phonons in 14 T
and to judge whether the phonons are nearly free from microscopic
scatterings at subKelvin temperatures. The phononic thermal
conductivity can be expressed by the kinetic formula $\kappa_{ph}
= \frac{1}{3}Cv_pl$,\cite{Berman} where $C = \beta T^3$ is the
phonon specific heat at low temperatures, $v_p$ is the average
velocity and $l$ is the mean free path of phonon. Here $\beta =
3.15 \times 10^{-3}$ J/K$^4$mol is obtained from the lattice
specific-heat data\cite{Yoshida1} and $v_p$ = 2440 m/s can be
estimated from $\beta$.\cite{Zhao_GFO, Tari} The obtained $l$ from
the 14 T $\kappa(T)$ data are compared with the averaged sample
width $W = 2\sqrt{A/\pi}$ = 0.993 mm,\cite{Berman, Sun_Comment}
where $A$ is the area of cross section. As shown in the inset to
Fig. 3, the ratio $l / W$ increases with lowering temperature and
becomes close to one at 0.3 K, which means that the boundary
scattering limit is nearly established at such low temperatures.
In other words, the heat transport at such low temperatures is
mainly contributed by the phonons, on which the microscopic
scatterings are very weak. It is known from the $H-T$ phase
diagram that 14-T field is strong enough to suppress the magnetic
ordering and fully polarize the spins, so the low-energy magnons
are hardly to be excited in such high field.\cite{Wang_HMO,
Zhao_GFO, Zhao_NCO} Therefore, it is naturally expected that the
phonon scattering by magnons is almost smeared out in 14 T. This
means that in zero and low fields, magnon excitations are mainly
playing a role of phonon scatterers.

\begin{figure}
\includegraphics[clip,width=6.0cm]{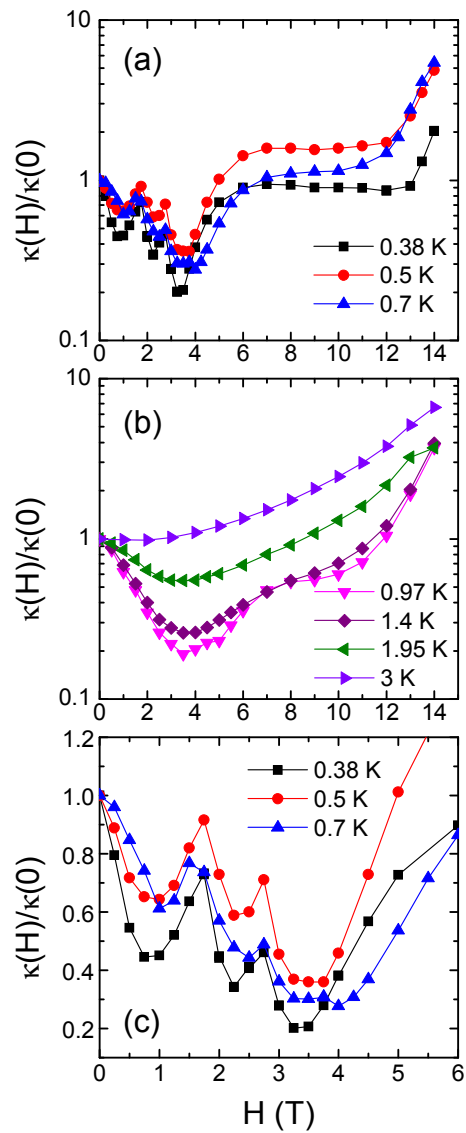}
\caption{(Color online) Magnetic-field dependencies of thermal
conductivity of MCCL crystal at low temperatures. The magnetic
fields are applied perpendicular to the $bc$ plane. (a) $T$ =
0.38--0.7 K. (b) $T$ = 0.97--3 K. (c) Zoom in of the low-field
plots for $T$ = 0.38--0.7 K.}
\end{figure}

To further clarify the roles of magnetic excitations and magnetic
phase transitions in the low-$T$ heat transport of MCCL, it is
useful to study the detailed magnetic-field dependence of
$\kappa$. Figures 4(a) and 4(b) show the $\kappa(H)$ isotherms at
temperatures below and above the zero-field $T_N$ = 0.9 K,
respectively. The qualitative behaviors of $\kappa(H)$ are
essentially the same for temperatures below $T_N$. At low-field
region, the $\kappa$ strongly decreases with $H$, accompanied with
three ``dips" at $\sim$ 1 T, 2.25 T, and 3.5 T. A ``plateau"-like
feature then appears at the intermediate field region, which is
terminated by a quick increase of $\kappa$ at high fields, with
the transition fields precisely coincided with the upper phase
boundary of FIMO. The strong increase of $\kappa$ at the
spin-polarized state can be clearly attributed to the weakening of
the phonon scattering by magnons because the number of the
low-energy magnons is quickly decreased.\cite{Wang_HMO, Zhao_GFO,
Zhao_NCO} This result can be compared to the observations of
$\kappa(T)$ in Fig. 3. As the $\kappa(H)$ curves show, even 14 T
field is not strong enough to completely suppress the magnetic
scattering on phonons. This is the reason that the 14 T
$\kappa(T)$ do not exhibit a precise $T^3$ dependence at subKelvin
temperatures.\cite{Sun_Comment} Nevertheless, in zero and low
fields, the magnetic scattering on phonons are significant.

The low-field $\kappa(H)$ behaviors are shown more clearly in Fig.
4(c). Apparently, the dip fields at $\sim$ 2.25 and 3.5 T
correspond precisely to the lower and upper critical fields of the
field-induced spin gapped state, respectively. The diplike
features at these two critical fields are attributable to the
phonon scattering by the strong critical spin fluctuations at the
phase transitions.\cite{Sun_DTN, Sologubenko5} The first dip at
$\sim$ 1 T locates in the low-field AF ordered state and is of
different origin. As can be seen in Fig. 2(a), the specific heat
data below 0.7 K do not show any anomaly or obvious field
dependence from 0 to 1.5 T. It is known that the spin-flop-like
transition of spin structure may not be detectable by the
specific-heat measurements. It is therefore very likely that the
1-T dip at very low temperatures is related to some kind of
spin-flop transition, which is also reasonable considering the
nearly temperature independence of the dip field.\cite{Spin_flop,
Wang_HMO, Zhao_GFO, Zhao_NCO} The spin-flop transition has been
indeed observed in an earlier magnetization
measurement,\cite{Inagaki1} although the transition field found at
$\sim$ 0.5 T is somewhat lower than that in the heat transport
data.

Another peculiar feature of the low-$T$ $\kappa(H)$ isotherms is a
``plateau" at the intermediate field region. At 0.38 K, for
example, the $\kappa$ is nearly field independent from 6 to 13 T.
With increasing temperature, the plateau becomes narrower and
disappears above 0.97 K. This field-independence phenomenon is
apparently in good agreement with the specific-heat data shown in
Fig. 2(c), which are also essentially independent on field above 6
T and at very low temperatures. One clear point is that the
plateau behavior shows up when the sample is in the FIMO state,
which indicates that the magnon spectrum at such low-$T$ phase
does not change strongly with applying field.

Above $T_N$, the three low-field ``dips" disappear; instead, the
$\kappa(H)$ curves show a broad valley-like behavior for $T$ =
0.97, 1.4 and 1.95 K, as shown in Fig. 4(b). In these curves, the
fields for the minimum $\kappa$ are all located at 3.5 T, which
indicates that it is not related to the lower transition field of
the FIMO phase. The $\kappa$ increase rather rapidly at higher
fields, demonstrating that the spin fluctuations are significantly
weakened. At even higher temperature of 3 K, the $\kappa$ shows a
monotonically increase with increasing field. Note that the
high-field-induced increase of $\kappa$ is actually happened in a
very broad temperature region, as also can be seen in Fig. 3.
Since the spontaneous or field-induced magnetic orders are not
relevant, the strong field dependence of $\kappa$ in this
temperature region is apparently related to the strong quantum
fluctuations of the low-dimensional spin systems. The data
indicates that the strong magnetic field tends to effectively
suppress the spin fluctuations, which can strongly scatter
phonons.

From above data and discussions, it is reasonably concluded that
there is no clear signature that the magnetic excitations in MCCL
have strong ability of transporting heat directly. This is rather
different from some well-studied low-dimensional spin systems,
like SrCuO$_2$, Sr$_2$CuO$_3$, CaCu$_2$O$_3$,
Sr$_{14}$Cu$_{24}$O$_{41}$, La$_2$CuO$_4$,
etc.,\cite{Sologubenko2, Kawamata, Hess2, Hess3, Sun_LCO,
Sun_PLCCO, Berggold} in which the magnetic excitations show a
remarkably strong heat conduction and they are hardly to be
affected by the laboratory magnetic field.\cite{Hess3, Brenig2}
The main reason is that those inorganic materials usually have
much larger exchange coupling, typically being of the order of
magnitude of 100 meV. Furthermore, the spin structure of MCCL is
more complex. For example, in the spontaneous AF phase only the FM
dimers ordered antiferromagnetically while the AF dimers are in
the $S$ = 0 ground state with strong quantum fluctuations.

It is also useful to note that the low-$T$ $\kappa(H)$ shows a
minimum at the field-induced quantum phase transition from the
spin-gapped state to the FIMO state. This is rather similar to
another FIMO material, Ba$_3$Mn$_2$O$_8$,\cite{Ke_BMO} in which
the FIMO can be discussed on the basis of the Bose-Einstein
condensation of magnons.\cite{BEC_Review} In contrast, another
magnon BEC compound, DTN,\cite{Sun_DTN} has shown strong ability
of magnetic heat transport along its spin-chain direction at the
phase transition from the gapped state to the field-induced AF
state.

\section{Summary}

The low-temperature heat transport in the quasi-one-dimensional
$S$ = 1/2 alternating antiferromagnetic-ferromagnetic chain
compound (CH$_3$)$_2$NH$_2$CuCl$_3$ is found to exhibit very
complicated temperature and magnetic-field dependencies. In zero
field, the strong spin fluctuations of this low-dimensional system
scatter phonons strongly in a broad temperature region from 0.3 to
30 K. The scattering is weakened when the spontaneous AF ordering
is formed below $T_N$, while it is enhanced when the AF order is
suppressed in magnetic fields. In higher magnetic fields, at the
phase transitions from the low-field N\'{e}el state to the
spin-gapped state ($\sim$ 2 T) and the field-induced magnetic
ordered state ($\sim$ 3.5 T), the thermal conductivity shows
diplike anomalies, which suggests the strong phonon scattering by
the critical fluctuations. In high fields, when the spins are
being polarized, the phonon conductivity is significantly enhanced
because of the weakening of spin fluctuations. Due to the complex
low-$T$ phase diagram and the multiple field-induced QPTs, it is
now difficult to perform the quantitative analysis on either the
temperature or the magnetic-field dependencies of $\kappa$.
Further knowledge about the magnetic spectra of this material are
necessary for deeper understanding of the heat transport
properties.

\begin{acknowledgements}

We thank W. Tao for technical assistance. This work was supported
by the Chinese Academy of Sciences, the National Natural Science
Foundation of China and the National Basic Research Program of
China (Grant Nos. 2009CB929502 and 2011CBA00111).

\end{acknowledgements}


\begin{thebibliography}{}

\bibitem{Sachdev}
S. Sachdev, Nature Phys. {\bf 4}, 173 (2008); and references
therein.

\bibitem{Balents}
L. Balents, Nature (London) {\bf 464}, 199 (2010).

\bibitem{BEC_Review}
For a review, see T. Giamarchi, C. R\"{u}egg, and O. Tchernyshyov,
Nature Phys. {\bf 4}, 198 (2008).

\bibitem{Brenig1}
F. Heidrich-Meisner, A. Honecker, and W. Brenig, Eur. Phys. J.
Special Topics {\bf 151}, 135 (2007).

\bibitem{Hess1}
C. Hess, Eur. Phys. J. Special Topics {\bf 151}, 73 (2007).

\bibitem{Sologubenko1}
A. V. Sologubenko, T. Lorenz, H. R. Ott, and A. Friemuth, J. Low.
Temp. Phys. {\bf 147}, 387 (2007).

\bibitem{Ando}
Y. Ando, J. Takeya, D. L. Sisson, S. G. Doettinger, I. Tanaka, R.
S. Feigelson, and A. Kapitulnik, Phys. Rev. B {\bf 58}, R2913
(1998).

\bibitem{Sun_DTN}
X. F. Sun, W. Tao, X. M. Wang, and C. Fan, Phys. Rev. Lett. {\bf
102}, 167202 (2009).

\bibitem{Sologubenko2}
A. V. Sologubenko, K. Giann\`{o}, H. R. Ott, A. Vietkine, and A.
Revcolevschi, Phys. Rev. B {\bf 64}, 054412 (2001).

\bibitem{Kawamata}
T. Kawamata, N. Takahashi, T. Adachi, T. Noji, K. Kudo, N.
Kobayashi, and Y. Koike, J. Phys. Soc. Jpn. {\bf 77}, 034607
(2008).

\bibitem{Hess2}
C. Hess, H. ElHaes, A. Waske, B. B\"{u}chner, C. Sekar, G.
Krabbes, F. Heidrich-Meisner, and W. Brenig, Phys. Rev. Lett. {\bf
98}, 027201 (2007).

\bibitem{Hess3}
C. Hess, C. Baumann, U. Ammerahl, B. B\"{u}chner, F.
Heidrich-Meisner, W. Brenig, and A. Revcolevschi, Phys. Rev. B
{\bf 64}, 184305 (2001).

\bibitem{Sologubenko3}
A. V. Sologubenko, K. Berggold, T. Lorenz, A. Rosch, E. Shimshoni,
M. D. Phillips, and M. M. Turnbull, Phys. Rev. lett. {\bf 98},
107201 (2007).

\bibitem{Sologubenko4}
A. V. Sologubenko, T. Lorenz, J. A. Mydosh, A. Rosch, K. C.
Shortsleeves, and M. M. Turnbull, Phys. Rev. lett. {\bf 100},
137202 (2008).

\bibitem{Sologubenko5}
A. V. Sologubenko, T. Lorenz, J. A. Mydosh, B. Thielemann, H. M.
R{\o}nnow, Ch. R\"{u}egg, and K. W. Kr\"{a}mer, Phys. Rev. B {\bf
80}, 220411(R) (2009).

\bibitem{Kohama_DTN}
Y. Kohama, A. V. Sologubenko, N. R. Dilley, V. S. Zapf, M. Jaime,
J. A. Mydosh, A. Paduan-Filho, K. A. Al-Hassanieh, P. Sengupta, S.
Gangadharaiah, A. L. Chernyshev, and C. D. Batista, Phys. Rev.
Lett. {\bf 106}, 037203 (2011).

\bibitem{Ajiro}
Y. Ajiro, K. Takeo, Y. Inagaki, T. Asano, A. Shimogai, M. Mito, T.
Kawae, K. Takeda, T. Sakon, H. Nojiri, M. Motokawa, Physica B {\bf
329-333}, 1008 (2003).

\bibitem{Willett}
R. D. Willett, B. Twamley, W. Montfrooij, G. E. Granroth, S. E.
Nagler, D. W. Hall, Ju-Hyun Park, B. C. Watson, M. W. Meisel, and
D. R. Talham, Inorg. Chem. {\bf 45}, 7689 (2006).

\bibitem{Inagaki1}
Y. Inagaki, A. Kobayashi, T. Asano, T. Sakon, H. Kitagawa, M.
Motokawa, and Y. Ajiro, J. Phys. Soc. Jpn. {\bf 74}, 2683 (2005).

\bibitem{Yoshida1}
Y. Yoshida, O. Wada, Y. Inagaki, T. Sakon, K. Takeo, T. Kawae, K.
Takeda, and Y. Ajiro, J. Phys. Soc. Jpn. {\bf 74}, 2917 (2005).

\bibitem{Furukawa}
Y. Furukawa, Y. Nishisaka, K. Kumagai, T. Asano, and Y. Inagaki,
J. Phys.: Conf. Ser. {\bf 51}, 87 (2006).

\bibitem{Stone1}
M. B. Stone, W. Tian, G. E. Granroth, M. D. Lumsden, J.-H. Chung,
D. G. Mandrusa, and S. E. Nagler, Physics B {\bf 385-386}, 438
(2006).

\bibitem{Stone2}
M. B. Stone, W. Tian, M. D. Lumsden, G. E. Granroth, D. Mandrus,
J.-H. Chung, N. Harrison, and S. E. Nagler, Phys. Rev. Lett. {\bf
99}, 087204 (2007).

\bibitem{Yoshida2}
Y. Yoshida, Y. Kitano, Y. Inagaki, T. Sakurai, M. Kimata, S.
Okubo, H. Ohta, K. Koyama, M. Motokawa, T. Asano, and Y. Ajiro, J.
Phys. Soc. Jpn. {\bf 76}, 113704 (2007).

\bibitem{Inagaki2}
Y. Inagaki, O. Wada, K. Ienaga, H. Morodomi, T. Kawae, Y. Yoshida,
T. Asano, Y. Frukawa, and Y. Ajiro, J. Phys.: Conf. Ser. {\bf
150}, 042067 (2009).

\bibitem{note}
It should be note that in some literature another description of
the crystal structure is used.\cite {Stone1,Stone2} It is also a
monoclinic symmetry at room temperature (space group $I2/a$) with
lattice constants $a$ = 11.97 \AA, $b$ = 8.63 \AA, $c$ = 14.34
\AA, and $\beta$ = 97.47$^\circ$.

\bibitem{Chen}
L. M. Chen, W. Tao, Z. Y. Zhao, Q. J. Li, W. P. Ke, X. M. Wang, X.
G. Liu, C. Fan, and X. F. Sun, J. Cryst. Growth {\bf 312}, 3243
(2010).

\bibitem{Wang_HMO}
X. M. Wang, C. Fan, Z. Y. Zhao, W. Tao, X. G. Liu, W. P. Ke, X.
Zhao, and X. F. Sun, Phys. Rev. B {\bf 82}, 094405 (2010).

\bibitem{Zhao_NCO}
Z. Y. Zhao, X. M. Wang, B. Ni, Q. J. Li, C. Fan, W. P. Ke, W. Tao,
L. M. Chen, X. Zhao, and X. F. Sun, Phys. Rev. B {\bf 83}, 174518
(2011).

\bibitem{Berman}
R. Berman, {\it Thermal Conduction in Solids} (Oxford University
Press, Oxford, 1976).

\bibitem{Sun_GBCO}
X. F. Sun, A. A. Taskin, X. Zhao, A. N. Lavrov, and Y. Ando, Phys.
Rev. B {\bf 77}, 054436 (2008).

\bibitem{Zhao_GFO}
Z. Y. Zhao, X. M. Wang, C. Fan, W. Tao, X. G. Liu, W. P.Ke, F. B.
Zhang, X. Zhao, and X. F. Sun, Phys. Rev. B {\bf 83}, 014414
(2011).

\bibitem{Tari}
A. Tari, {\it Specific Heat of Matter at Low Temperatures}
(Imperial College Press, 2003).

\bibitem{Sun_Comment}
X. F. Sun and Y. Ando, Phys. Rev. B {\bf 79}, 176501 (2009).

\bibitem{Spin_flop}
J. A. H. M. Buys and W. J. M. de Jonge, Phys. Rev. B {\bf 25},
1322 (1982); G. S. Dixon, {\it ibid.} {\bf 21}, 2851 (1980).

\bibitem{Sun_LCO}
X. F. Sun, J. Takeya, S. Komiya, and Y. Ando, Phys. Rev. B {\bf
67}, 104503 (2003).

\bibitem{Sun_PLCCO}
X. F. Sun, Y. Kurita, T. Suzuki, S. Komiya, and Y. Ando, Phys.
Rev. Lett. {\bf 92}, 047001 (2004).

\bibitem{Berggold}
K. Berggold, T. Lorenz, J. Baier, M. Kriener, D. Senff, H. Roth,
A. Severing, H. Hartmann, A. Freimuth, S. Barilo, and F. Nakamura,
Phys. Rev. B {\bf 73}, 104430 (2006).

\bibitem{Brenig2}
F. Heidrich-Meisner, A. Honecker, and W. Brenig, Phys. Rev. B {\bf
71}, 184415 (2005).

\bibitem{Ke_BMO}
W. P. Ke, X. M. Wang, C. Fan, Z. Y. Zhao, X. G. Liu, L. M. Chen,
Q. J. Li, X. Zhao, and X. F. Sun, Phys. Rev. B {\bf 84}, 094440
(2011).


\end{thebibliography}
\end{document}